\documentclass{aa}
\usepackage{graphicx}
\usepackage{txfonts}

\def\apgt{\ {\raise-.5ex\hbox{$\buildrel>\over\sim$}}\ }
\def\aplt{\ {\raise-.5ex\hbox{$\buildrel<\over\sim$}}\ }

\newcommand{\msun}{\mbox{${\rm M}_\odot$}}

\newcommand{\kms}{\mbox{${\rm km~s}^{-1}$}}

\begin{document}
%\headnote{Letters to the editor}

%\title{A catalogue of low-mass X-ray binaries}
\title{A catalogue of low-mass X-ray binaries in the Galaxy, LMC, and SMC  (Fourth edition)
\thanks {Table 1 is also available in electronically form at the CDS via anonymous ftp
(130.79.128.5) or via http://cdsweb.u-strasbg.fr/cgi-bin/qcat?J/A+A/}
 }
\author{Q.Z. Liu\inst{1, 2},
                J. van Paradijs\inst{2},
and E.P.J. van den Heuvel\inst{2}}

\offprints{Q.Z. Liu}
%\offprints{Q.Z. Liu (email: qzliu@astro.uva.nl)}

\institute{Purple Mountain Observatory, Chinese Academy of Sciences, Nanjing 210008, P.R. China\\
           \email{qzliu@pmo.ac.cn}
 \and
      Astronomical Institute "Anton Pannekoek", University of Amsterdam, Kruislaan 403,
           1098 SJ Amsterdam, The Netherlands}

\date{Received date/ Accepted date}
%\markboth{Q.Z. Liu et al.:\,\, A catalogue of low-mass X-ray binaries}{}

 \setcounter{page}{1}

\abstract  % context heading (optional)
  {}
  {The aim of this catalogue is to provide the reader with some basic
information on the X-ray sources and their counterparts in other wavelength ranges ($\gamma$-rays, UV, optical, IR, and
radio). Some sources, however, are only tentatively identified as low-mass X-ray binaries on the basis of their X-ray
properties similar to the known low-mass X-ray binaries. Further identification in other wavelength bands is needed to
finally determine the nature of these sources. In cases where there is some doubt about the low-mass nature of the
X-ray binary this is mentioned. }
{Literature published before 1 October 2006 has, as far as possible, been taken into account.}
{We present a new edition of the catalogue of the low-mass X-ray binaries in the Galaxy and the Magellanic Clouds. The
catalogue contains source name(s), coordinates, finding chart, X-ray luminosity, system parameters, and stellar
parameters of the components and other characteristic properties of 187 low-mass X-ray binaries, together with a
comprehensive selection of the relevant literature. }
 {}


 \maketitle

\authorrunning{Q.Z. Liu, J. van Paradijs and E.P.J. van den Heuvel}
\titlerunning{A catalogue of low-mass X-ray binaries}

\section{Introduction}
%LMXB number about 200 (in't Zand et al. 2005). Six LMXBs had persistent emission levels below the detection limit (Cornelisse et al. 2002, A\&A).
 An X-ray binary contains either a neutron star (NS) or a black hole (BH) accreting
material from a companion star. Thus, white dwarf (WD) systems are not included among the X-ray binaries. X-ray
binaries can be further divided into two different classes, regardless the nature of the compact object, according to
the mass of the companion star: high-mass X-ray binaries (HMXB) and low-mass X-ray binaries (LMXB). In a previous paper
we presented an up-to-date catalogue of high-mass X-ray binaries (Liu, van Paradijs \& van den Heuvel 2006).

The secondary of LMXB systems is a low-mass (in general $M \leq 1 \msun $) star, which transfers matter by Roche-lobe
overflow. Among the low-mass companion stars we find white dwarfs, late-type main-sequence stars, A-type stars and
F-G-type sub-giants. The last-mentioned category of companion stars may well be the mass-transfer remnants of stars
that originally were of intermediate mass ($M\sim$1.5 to 4 $\msun$) as has been suggested for Cygnus X-2 (cf.
Podsiadlowski \& Rappaport 2000; Tauris, van den Heuvel \& Savonije 2000). The optical counterparts of LMXBs are
intrinsically faint objects. The spectra of most of them show a few characteristic emission lines superposed on a
rather flat continuum. The optical continuum of LMXBs is dominated by the emission from an accretion disk around the
compact star, which is predominantly the result of reprocessing of a fraction of the X-rays into optical photons in the
disk. The contribution from the secondary is generally negligible. On occasions, however, the presence of the secondary
can be discerned in the spectrum (or colors) of the LMXBs. This particularly is the case for systems with donors that
are or started out as intermediate mass stars, such as Her X-1 and Cyg X-2. For a full understanding of LMXBs in
various aspects one can refer to the book by Lewin \& van der Klis (2006).

The classification as LMXB is mainly based on the spectra obtained from an optical identification, and/or on the mass
function from X-ray pulse arrival time measurements. If neither is available, a classification may be inferred based on
the similarity of the X-ray properties to other identified systems. An unidentified system is classified as a LMXB
containing a neutron star if one or more of the following properties are observed:

\begin{itemize}
\item[$\bullet$] type I X-ray bursts (which to date have only been seen from neutron stars in LMXBs);

\item[$\bullet$] a 1-10 keV soft spectrum with a characteristic temperature of 5-10 keV and/or

\item[$\bullet$] an orbital period is less than about 12 hr.
\end{itemize}

Six years after the publication of the previous (3rd) edition (Liu, van Paradijs \& van den Heuvel 2001), the amount of
new literature and the number of new objects to be included have again grown so much that it seems worthwhile to
publish an updated catalogue. This new catalogue contains 187 sources, 44 new low-mass X-ray binaries in addition to
the 143 sources listed in the previous catalogue. We briefly recall some of the developments that, over the past six
years, have had (and still have) a major impact on this catalogue.

Due to the much increased sensitivity and spatial resolution achievable with the {\it Chandra} and the {\it XMM-Newton}
X-ray observatories, as well as with the unparalleled resolution of the Hubble Space Telescope and large ground-based
radio telescopes, more accurate positions of X-ray binaries have been determined, resulting in the unambiguous
discovery of the optical and/or IR counterpart to some X-ray sources. Moreover, the number of LMXBs in external
galaxies is also rapidly increasing, e.g., the X-ray binaries in NGC 720 (Jeltema et al. 2003) and NGC 1399 (Angelini
et al. 2001). Barnard et al. (2003) even discovered a Z source low-mass X-ray binary, RX J0042.6+4115, in M 31. Most of
the ultra-luminous X-ray sources in elliptical galaxies (Liu \& Mirabel 2005) are believed to be LMXBs with a black
hole. It has been proposed that the collective X-ray luminosity of low-mass X-ray binaries can be used as a stellar
mass indicator for the host galaxy (Gilfanov 2004).

Prior to {\it Chandra} and {\it XMM}, there were 12 bright cluster LMXBs (now 13, all thought to have neutron stars as
primaries) and 57 faint X-ray sources known in the Galactic globular cluster systems (Verbunt 2001). A few of the
latter had been identified with cataclysmic variables, and some were thought to be quiescent LMXBs (qLMXBs) containing
neutron stars (Verbunt et al. 1984). Much more quiescent low-mass X-ray binaries have been identified in several
globular clusters using {\it Chandra} or {\it XMM} X-ray observations (e.g. NGC 6440, Heinke et al. 2003). These
qLMXBs, however, have not been included in this catalogue.

Since its launch in 2002, {\it INTEGRAL} has been revealing hard X-ray sources that were not easily detected in earlier
soft X-ray (typically $\leq$10 keV) observations (Winkler et al. 2003). Hard X-rays are not easily absorbed by matter
and thus are highly penetrating. Such radiation is, therefore, ideal for probing high-energy emitting sources in dense
regions. {\it INTEGRAL} has been performing a regular survey of the Galactic plane and a deep exposure of the Galactic
Center as part of its Core Program. Bird et al. (2007) have recently published an updated catalogue of the {\it
INTEGRAL} X-ray sources. The catalogue contains 421 point X-ray sources, some of which are included in our catalogue.
For the details on new {\it INTEGRAL} sources please refer to the web page of Jerome Rodriguez
(http://isdc.unige.ch/$^\sim$rodrigue/html/igrsources.html).

The aim of our catalogue is to provide some basic information on the X-ray sources and their counterparts, as well as
the binary properties of the system in question, and easy access to the recent literature. No attempt has been made to
compile complete reference lists. Much effort has been made to avoid errors and to keep the information up to date.
Nevertheless, the authors are well aware that this edition too may contain errors and may be incomplete.
%It is certainly incomplete with respect to the references quoted.

%All the tabular material contained in this catalogue is published in electronic form only. It
%is available in electronic form at the CDS via anonymous ftp to cdsarc.u-strasbg.fr
%(130.79.128.5) or via http://cdsweb.u-strasbg.fr/cgi-bin/qcat?J/A+A/404/301.

%In addition to the electronic version we provide postscript files for a printable stand alone
%version at astro-ph.

\section{Description of the table}

Table 1 lists the 187 LMXBs. The format of Table 1 is similar to the previous edition (Liu et al. 2001), of which the
present catalogue is meant to be an update. In the table the sources are ordered according to the right ascension in
the source names; part of the (mainly numerical) information on a source is arranged in six columns, below which
additional information is provided for each source in the form of key words with reference numbers [in square
brackets]. When a result is unreliable, a colon (:) or a question mark (?) will follow the adopted entry. The columns
have been arranged as follows.

In Column 1 the first line contains the source name, which contains an indication on its sky location, according to the
conventional source nomenclature of space experiments in which the source was detected, hhmm$\pm$ddd or
hhmm.m$\pm$ddmm. Here hh, mm, and ss indicate the hours, minutes and seconds of right ascension, ddd the declination in
units of 0.1 degree (in a small number of cases, the coordinates shown in the name are given with more, or fewer,
digits). The prefix J indicates a name based on J2000 coordinates. Otherwise, 1950 coordinates were used in the name.
An alternative source name is given in the second line. In the third line of Column 1, the source types are indicated
with a letter code, as follows:
\begin{itemize}
 \item[$\bullet$] A: known atoll source (25);

     The majority of systems in the catalogue are unclassified. However, advances in our
     knowledge and understanding of the properties of the neutron-star X-ray binaries indicate that the majority of the
     unclassified systems are likely to be Atoll-like, although a small subclass of lower-luminosity sources is possible
     (Fender \& Hendry 2000).
 \item[$\bullet$] B: X-ray burst source (84);

     There are a dozen of superbursts detected from ten sources: one from 4U 1735-44, Ser X-1, KS 1731-260,
     4U 1820-303, GX 3+1, 4U 0614+091, 4U 1608-522, 4U 1254, SLX 1735-269 and three from 4U 1636-536.
     It is thought that unstable carbon burning (Woosley \& Taam 1976; Strohmayer \& Brown 2002) in a heavy-element
     ocean (Cumming \& Bildsten 2001), possibly combined with photo-disintegration-triggered nuclear energy
     release (Schatz et al. 2003), is responsible for most superbursts.
 \item[$\bullet$] D: "dipping" low-mass X-ray binary (22);
 \item[$\bullet$] E: eclipsing or partially eclipsing low-mass X-ray binary (13);
 \item[$\bullet$] G: globular-cluster X-ray source (14);
 \item[$\bullet$] M: microquasar (14);
 \item[$\bullet$] P: X-ray pulsar (12);

     These sources include the 7 currently known accretion-driven millisecond X-ray pulsars (IGR J00291+5934,
     XTE J0929-314, XTE J1751-305, XTE J1807-294, SAX J1808.4-3658, XTE J1814-338, and HETE J1900.1-2455).

 \item[$\bullet$] R: radio loud X-ray binary (56);
 \item[$\bullet$] T: transient X-ray source (103);
 \item[$\bullet$] U: ultra-soft X-ray spectrum (18);

     These sources include black-hole candidates; some 'extreme ultra-soft' (EUS) sources may be white dwarfs
     on whose surface steady nuclear burning takes place.
 \item[$\bullet$] Z: Z-type source (8), including the possible new Z source, XTE J1701-462 (Homan et al. 2006).
\end{itemize}

Column 2 contains the right ascension (RA) and declination (DEC) of the source for equinox J2000.0 in the first two
lines. RA is given as hhmmss.s to an accuracy of 0.1 s, DEC is given in $^\circ$ ' ", to an accuracy of 1" (in a small
number of cases, the coordinates are given with more, or fewer, digits). The third line gives the galactic longitude
and latitude to an accuracy of 0.1$^\circ$ (except for sources close to the galactic center, where these coordinates
are given to 0.01$^\circ$). A reference for the source position is specified in the table after $'Pos.'$. In the
parentheses following the $'pos.'$, we provide some information on the type of observation from which the source
position has been derived. The following abbreviations are used: o, optical; x, X-ray; r, radio; IR, infrared.
Following the type of observation, we give an indication of the accuracy of this position, in the form of equivalent
(90 percent confidence level) error radii, but in several cases this can only be considered an approximation (e.g. when
the error box is not circular). %When no accuracy is quoted, it is about one arcsecond or better.

The first and second lines of Column 3 give the names of the optical counterpart to an X-ray source. The third line
contains a reference, in which the finding chart to the X-ray source can be found. An asterisk followed by a number or
letter refers to the star number used in the finding chart; "star" refers to a star in the finding chart that has not
been assigned a number or letter. If there is only a reference but nothing else specified in the first two lines, that
means there is no optical counterpart to be found. Many optical counterparts have been indicated with a variable-star
name, as given in the $General~Catalogue~of~Variable~Stars$ and in recent name lists of variable stars as published
regularly in the $IAU~Information~Bulletin~on~Variable~Stars$, or a number in a well-known catalogue (e.g. HD, SAO).
For X-ray sources in globular clusters, the cluster name is given, in addition to the name of a stellar optical
counterpart.

The fourth column contains some photometric information on the optical counterpart. In the first line, the apparent
visual magnitude, $V$, and the color indices $B-V$, and $U-B$, are listed. The second line contains the estimate of the
interstellar reddening, $E_{B-V}$. Some of them are derived from the best-fit column density $N_H$ (in $cm^{-2}$) of
their X-ray spectrum, through the following relation (Predehl \& Schmitt 1995),
\begin{eqnarray*}
  A_V    & = &  \frac{N_H}{1.79 \times 10^{21}}   \\
 E(B-V)  & = &  A_V/3.2.
 \end{eqnarray*}

In Column 5, the maximum X-ray flux, or the range of observed X-ray fluxes (2-10 keV, unless otherwise indicated), is
given, in units of
%1$\mu$Jy = 10$^{-29}$~erg~cm$^{-2}$~s$^{-1}$~Hz$^{-1}$\\
% ~~~~~   = 2.4$\times$ 10$^{-12}$~erg~cm$^{-2}$~s$^{-1}$~keV$^{-1}$.
%        = 2.2$\times$ 10$^{-11}$~erg~cm$^{-2}$~s$^{-1}$ at 2-10~keV
%        = 2.6$\times$ 10$^{35} (d/10kpc)^2$ ~erg~s$^{-1}$ at 2-10 keV
%%% ½ñºóÓÃ
\begin{eqnarray*}
1\mu Jy & = & 10^{-29}~ erg~ cm^{-2}~ s^{-1} Hz^{-1}\\
        & = & 2.4\times 10^{-12}~ erg~  cm^{-2}~  s^{-1}~ keV^{-1}\\
%        & = & 2.2\times 10^{-11}~erg~cm^{-2}~s^{-1}~ at~ 2-10~keV\\
%        & = & 2.6\times 10^{35} (\frac{d}{10kpc})^2 ~erg~s^{-1}~ at~ 2-10 keV.
\end{eqnarray*}
% ISGRI 20-60 keV (1 Crab $\sim$ 160 counts/s)               60-150 keV (1Crab $\sim$ 40 counts/s)
% JEM-X 3-10  keV (1 Crab on axis $\sim$ 105-115 counts/s)   10-25  keV (1 Crab on axis $\sim$ 45-50 counts/s)
% RXTE PCA: Note that 1 Crab (2-10 keV) = 2.42 $\times$ 10$^{-8}$ ergs/s/cm$^2$ = 2.27 counts/s/ PCU.
% RXTE ASM: 3 counts/s, equivalent to roughly 1.2$\times10^{-9}$ ergs/cm$^2$/s in the 2¨C20 keV band (Kaaret, P. et al.
%2005, ApJ 638)..... assuming that 1 Crab = 2.975$\times$ 10$^{-8}$ ergs/s/cm$^2$ (1.5 - 12keV) (Augusteijn, T. et al.
%2001, A\&A, 375). 1 Crab = 74 ASM counts/s (Tomsick et al. 05, ApJ, 630)
% BeppoSAX: 100 mCrab$\sim$1.8$\times$x10$^{37}$ (9 kpc) (2-10keV)
% Chandra :   1 mCrab$\sim$3  $\times$x10$^{35}$ (9 kpc) (0.5-10 keV)
% $A_B$=1.32$A_V$, $A_R$=0.81$A_V$

The first line in Column 6 gives the orbital period in hours (in a small number of cases with very wide orbit, the unit
of orbital period is in days, indicated with a 'd' after the number). The second line contains the pulse period for
X-ray pulsars, in seconds. The third line contains a reference in which the orbital and/or pulse periods were detected.

\section{Conclusions and remarks}
The current version of this catalogue provides tabulated data and references for 187 objects, including 44 newly
discovered LMXBs (2 previously listed in our high-mass X-ray binary catalogue), as well as 143 $``$old" ones listed in
the previous catalogue. Compared with the 3rd edition, the number of LMXBs in the Galaxy listed has increased by
$\sim$30\%. Among the 187 LMXB candidates, we find 3 symbiotic LMXBs with M-type giant companion (4U 1700+24, 4U
1954+319, and GX 1+4), 15 known and suspected ultra-compact X-ray binaries (UCXBs) (4U 0614+091, 4U 0919-54, 1A
1246-588, 4U 1543-624, 4U 1626-67, XTE J1709-267, EXO 1745-248, 4U 1812-12, 4U 1820-30, 4U 1823-00, XB 1832-330, 4U
1850-087, 4U 1905+000, 4U 1916-05, and CXO J212958.1+121002), and 14 sources with millisecond burst oscillations. UCXBs
are those systems with orbital periods $<80$ min, which is the minimum period for LMXBs with hydrogen-rich main
sequence donors. In these ultracompact binaries the mass donor must be a non-degenerate hydrogen-deficient star or a
white dwarf (e.g. Verbunt \& van den Heuvel 1995). In addition to the 7 millisecond X-ray pulsars, there are 14 X-ray
sources with millisecond burst oscillations. It is certain that the sources are neutron stars with millisecond spin
periods, since Chakrabarty et al. (2003) found compelling evidence that burst oscillation is the same as the X-ray
pulsation.

We wish to emphasize here that some sources listed in this catalogue are still uncertain. They need to be regarded with
caution in view of all the further work needed. Some sources are tentatively classified as low-mass X-ray binaries due
to the similarity of the X-ray properties to other identified systems. No counterpart at other bands has been found.

We would like to make some remarks on several individual sources. Both 2S 1803-245 and XTE J1806-246 have been listed
in the 3rd edition. Marshall et al. (1998), reported that the bright, soft X-ray transient, XTE J1806-246, is at a
position consistent with that of 2S 1803-245. Therefore, in the new version of the catalogue, they are regarded as the
same source. Similarly, the BeppoSAX source, SAX J1748.9-2021, is the only one in the Uhuru error boxes of 4U 1745-203
(in't Zand et al. 1999), also regarded as the identical source.

The following six X-ray sources are no longer listed in this LMXB catalogue. Three previous LMXBs, MXB 0656$-$072, SAX
J1819.3$-$2525/V4641 Sgr, and GRS 1736-297, have been listed in the HMXB catalogue, because of their HMXB properties
(Liu et al. 2006). In SIMBAD databases, SAX J0835.9+5118 is an X-ray flash source, GRB 990520, instead of an X-ray
burster, thus excluded from the LMXB catalogue. For the other two previously proposed low-mass X-ray binaries, GRS
1734-292 and GT 2318+620, new optical and radio observations provide substantial evidences to rule out their galactic
origin and point towards identification as active extragalactic sources (Marti et al. 1998; Paredes et al. 2004).

The following two X-ray sources, 4U 1954+319 and 1A 1246-588, were previously listed as HMXBs but are now in this
catalogue. The peculiar galactic X-ray source, 4U 1954+319, has been listed in the previous editions as an HMXB.
Masetti et al. (2006), however, found that the suspected field M-type giant star is indeed the counterpart of the X-ray
source, based on the Chandra Observation. They suggest that 4U 1954+319 is a wide-orbit LMXB, most likely a neutron
star, accreting from the wind of an M-type giant.

1A 1246-588 was previously included in HMXB catalogue. However, RXTE observations revealed an unusual burst in the
source (Levine et al. 2006). Bassa et al. (2006) recently reported an optical identification of 1A 1246-588 which
suggests that the system is an ultracompact X-ray binary (thus a LMXB) and they note that type I bursts have been seen
from this system in BeppoSAX and RXTE data.

%XMM J174457-2850.3 is a faint X-ray transient that has not been classified yet and that was found at a 2-10 keV flux of
%approximately 9E-11 erg/s/cm$^2$ (assuming an absorbed power-law spectral shape with a Galactic column density of
%5.9E22 cm$^{-2}$ and a power-law index of 1.0; see Sakano et al. 2005).
%Yamauchi, S. 2005PASJ...57..465: An Ultrasoft Transient X-Ray Source in the Norma Region Discovered with Ginga. source
%H/1RXS J160807.7-504149?
%IGR J2018+4043 ATel 788, AGN or XB

\acknowledgements  We wish to thank Marc Rib\'o, Ada Paizis, Fraser Lewis, Lara Sidoli and an anonymous referee for
their carefully reading the manuscript and useful comments. We also thank Marc Rib\'o for providing us with information
on radio loud X-ray sources and Ada Paizis for access to her results prior to publication. This research has made use
of the SIMBAD data base, operated at CDS, Strasbourg, France, and NASA's Astrophysics Data System (ADS). QZL is
partially supported by the Royal Science Foundation of The Netherlands, the National Natural Science Foundation of
China under Grants 10673032 and 10433030.

%, Deepto Chakrabarty, Nicola Masetti, Ignacio Negueruela, Sylvain Chaty

%Noted added in proof: Rudy Wijnands informed us that we had missed

%\end{document}

%HE 0532-4503 ( 2006, ATEL 937: Swift observation of HE 0532-4503  E. M. Cackett, J. M. Miller)

\clearpage \begin{table*}[t] \caption{Low-mass X-ray binaries}
\begin{tabular}{p{3.8cm}p{1.9cm}p{2.3cm}p{3.5cm}p{2.0cm}p{2.0cm}}
\noalign{\smallskip}
%\multicolumn{6}{p{17.6cm}}{Table 1.  (continued) }\\
\hline \noalign{\smallskip}
Source     & RA(J2000)         & Opt. Ctp.  & V, B-V, U-B  & $F_{x} (\mu Jy)$  & P$_{orb}$(hr)    \\
Name(s)    & DEC(J2000)        &            & E$_{B-V}$    & (2-10 keV)        & P$_{pulse}$(s)   \\
type       & $l^{II}, b^{II}$  & [FC]       &              &                   &                  \\

 \noalign{\smallskip}\hline \noalign{\smallskip}
IGR J00291+5934  & 00 29 03.06 & star   &   & 52            & 2.46    \\
                 & +59 34 19.0 &        &   & (2.5--25 keV) & 0.00167 \\
T, P, R          & 120.1, -3.2 & [439]  &   & [461]         & [461]  \\
\\
\multicolumn{6}{p{17.6cm}}{$Pos$.(o 0.5"): [439]; X/$\gamma$-ray transient: [1397]; X-rays: [1397, 1188, 398]; very low
mass-donor: [761]; distance 2.6-3.6 kpc: [761]; optical counterpart with R$\sim$ 17.4 mag.: [439]; optical light curve:
[101]; weak He II and H$\alpha$ emission: [1322]; radio counterpart: [1250]; orbit and spin parameters: [461];
very similar to SAX J1808.4-3658: [461]. %B??????
}\\
\hline

\end{tabular}
\end{table*}

\clearpage \begin{table*}[t]
\begin{tabular}{p{3.8cm}p{1.9cm}p{2.3cm}p{3.5cm}p{2.0cm}p{2.0cm}}
\noalign{\smallskip}
\multicolumn{6}{p{17.6cm}}{Table 1.  (continued) }\\
\hline \noalign{\smallskip}
Source     & RA(J2000)         & Opt. Ctp.  & V, B-V, U-B  & $F_{x} (\mu Jy)$  & P$_{orb}$(hr)    \\
Name(s)    & DEC(J2000)        &            & E$_{B-V}$    & (2-10 keV)        & P$_{pulse}$(s)   \\
type       & $l^{II}, b^{II}$  & [FC]       &              &                   &                  \\

\noalign{\smallskip} \hline \noalign{\smallskip}
4U 0614+091    & 06 17 07.4       & V1055 Ori        & 18.5, 0.3, -0.5    & 50             &    \\
               & +09 08 13.6      &                  & 0.3-0.64           &                &    \\
A, B, R?       & 200.9, -3.4      & [368, 513]       & [328, 767, 1143]   & [1685, 1752]   &    \\
\\
\hline
\end{tabular}
\end{table*}

\clearpage \begin{table*}[t]
\begin{tabular}{p{3.8cm}p{1.9cm}p{2.3cm}p{3.5cm}p{2.0cm}p{2.0cm}}
\noalign{\smallskip}
\multicolumn{6}{p{17.6cm}}{Table 1.  (continued) }\\
\hline \noalign{\smallskip}
Source     & RA(J2000)         & Opt. Ctp.  & V, B-V, U-B  & $F_{x} (\mu Jy)$  & P$_{orb}$(hr)  \\
Name(s)    & DEC(J2000)        &            & E$_{B-V}$    & (2-10 keV)        & P$_{pulse}$(s) \\
type       & $l^{II}, b^{II}$  & [FC]       &              &                   &                \\

\noalign{\smallskip}\hline \noalign{\smallskip}
XTE J0929-314 & 09 29 20.19 & star   & 18.63, -0.13,  & 36     & 0.73         \\
              & -31 23 03.2 &        & 0.13           &        & 0.0054       \\
T, P, R       & 260.1, 14.2 & [489]  & [489]          & [457]  & [457, 1294] \\
\\
\hline
\end{tabular}
\end{table*}

\clearpage \begin{table*}[t]
\begin{tabular}{p{3.8cm}p{1.9cm}p{2.3cm}p{3.5cm}p{2.0cm}p{2.0cm}}
\noalign{\smallskip}
\multicolumn{6}{p{17.6cm}}{Table 1.  (continued) }\\
\hline \noalign{\smallskip}
Source     & RA(J2000)         & Opt. Ctp.  & V, B-V, U-B  & $F_{x} (\mu Jy)$  & P$_{orb}$(hr)    \\
Name(s)    & DEC(J2000)        &            & E$_{B-V}$    & (2-10 keV)        & P$_{pulse}$(s)   \\
type       & $l^{II}, b^{II}$  & [FC]       &              &                   &                  \\

\noalign{\smallskip} \hline \noalign{\smallskip}
4U 1323-62  & 13 26 36.31   & star    &           & 10            & 2.93   \\
            & -62 08 09.9   &         & $\sim$5:  &               &        \\
B, D        & 307.0, +0.46  & [1430]  & [1202]    & [462, 1685]   & [1202] \\
\\
\hline
\end{tabular}
\end{table*}

\clearpage \begin{table*}[t]
\begin{tabular}{p{3.8cm}p{1.9cm}p{2.3cm}p{3.5cm}p{2.0cm}p{2.0cm}}
\noalign{\smallskip}
\multicolumn{6}{p{17.6cm}}{Table 1.  (continued) }\\
\hline \noalign{\smallskip}
Source     & RA(J2000)         & Opt. Ctp.  & V, B-V, U-B  & $F_{x} (\mu Jy)$  & P$_{orb}$(hr)    \\
Name(s)    & DEC(J2000)        &            & E$_{B-V}$    & (2-10 keV)        & P$_{pulse}$(s)   \\
type       & $l^{II}, b^{II}$  & [FC]       &              &                   &                  \\

\noalign{\smallskip} \hline \noalign{\smallskip}
4U 1543-47  & 15 47 08.6     & IL Lup      & 14.9, 0.6, - -     & $<$1-15000    & 26.8    \\
            & -47 40 10      &             & 0.5                &               &          \\
T, U, R     & 330.9, +54     & [183]       & [1225, 1178]       & [128, 808]    & [1178]    \\
\\
\hline
\end{tabular}
\end{table*}

\clearpage \begin{table*}[t]
\begin{tabular}{p{3.8cm}p{1.9cm}p{2.3cm}p{3.5cm}p{2.0cm}p{2.0cm}}
\noalign{\smallskip}
\multicolumn{6}{p{17.6cm}}{Table 1.  (continued) }\\
\hline \noalign{\smallskip}
Source     & RA(J2000)         & Opt. Ctp.  & V, B-V, U-B  & $F_{x} (\mu Jy)$  & P$_{orb}$(hr)    \\
Name(s)    & DEC(J2000)        &            & E$_{B-V}$    & (2-10 keV)        & P$_{pulse}$(s)   \\
type       & $l^{II}, b^{II}$  & [FC]       &              &                   &                  \\

\noalign{\smallskip}\hline \noalign{\smallskip}
1E 161348-5055.1  & 16 17 36.26   &  &   & 1            & 6.67   \\
                  & -51 02 25.0   &  &   & (0.5-2 keV)  &        \\
                  & 332.4, -0.37  &  &   & [347]        & [347]  \\
\\
\hline

\end{tabular}
\end{table*}

\clearpage \begin{table*}[t]
\begin{tabular}{p{3.8cm}p{1.9cm}p{2.3cm}p{3.5cm}p{2.0cm}p{2.0cm}}
\noalign{\smallskip}
\multicolumn{6}{p{17.6cm}}{Table 1.  (continued) }\\
\hline \noalign{\smallskip}
Source     & RA(J2000)         & Opt. Ctp.  & V, B-V, U-B  & $F_{x} (\mu Jy)$  & P$_{orb}$(hr)    \\
Name(s)    & DEC(J2000)        &            & E$_{B-V}$    & (2-10 keV)        & P$_{pulse}$(s)   \\
type       & $l^{II}, b^{II}$  & [FC]       &              &                   &                  \\

\noalign{\smallskip} \hline \noalign{\smallskip}
4U 1642-45   & 16 45 47.7     &            &             & 500     &   \\
GX 340+0     & -45 36 39.5    &            & 12:         &         &   \\
Z, R         & 339.6, -0.1    & [11]       & [607]       & [1685]  &   \\
\\
\hline

\end{tabular}
\end{table*}

\clearpage \begin{table*}[t]
\begin{tabular}{p{3.8cm}p{1.9cm}p{2.3cm}p{3.5cm}p{2.0cm}p{2.0cm}}
\noalign{\smallskip}
\multicolumn{6}{p{17.6cm}}{Table 1.  (continued) }\\
\hline \noalign{\smallskip}
Source     & RA(J2000)             & Opt. Ctp.   & V, B-V, U-B  & $F_{x} (\mu Jy)$  & P$_{orb}$(hr)   \\
Name(s)    & DEC(J2000)            &             & E$_{B-V}$    & (2-10 keV)        & P$_{pulse}$(s)  \\
type       & $l^{II}, b^{II}$      & [FC]        &              &                   &                 \\

\noalign{\smallskip} \hline \noalign{\smallskip}
4U 1659-487       & 17 02 49.5      & V821 Ara    & 15.5, 0.8, -0.1    & 1.5-900       & 42.14      \\
GX 339-4          & -48 47 23       & $^*$V       & 1.1                &               &            \\
T, U, M, R        & 338.9, -4.3     & [153, 374]  & [938, 146]         & [696, 1086]   & [683] \\
\\
\hline
\end{tabular}
\end{table*}

\clearpage \begin{table*}[t]
\begin{tabular}{p{3.8cm}p{1.9cm}p{2.3cm}p{3.5cm}p{2.0cm}p{2.0cm}}
\noalign{\smallskip}
\multicolumn{6}{p{17.6cm}}{Table 1.  (continued) }\\
\hline \noalign{\smallskip}
Source     & RA(J2000)             & Opt. Ctp.   & V, B-V, U-B  & $F_{x} (\mu Jy)$  & P$_{orb}$(hr)   \\
Name(s)    & DEC(J2000)            &             & E$_{B-V}$    & (2-10 keV)        & P$_{pulse}$(s)  \\
type       & $l^{II}, b^{II}$      & [FC]        &              &                   &                 \\

\noalign{\smallskip} \hline \noalign{\smallskip}
4U 1705-44      & 17 08 54.47    &           &             & 10-280       &   \\
                & -44 06 07.35   &           & 2:          &              &   \\
B, A, R         & 343.3, -2.3    & [1277]    & [863]       & [863, 1685]  &   \\
\\
\hline
\end{tabular}
\end{table*}

\clearpage \begin{table*}[t]
\begin{tabular}{p{3.8cm}p{1.9cm}p{2.3cm}p{3.5cm}p{2.0cm}p{2.0cm}}
\noalign{\smallskip}
\multicolumn{6}{p{17.6cm}}{Table 1.  (continued) }\\
\hline \noalign{\smallskip}
Source     & RA(J2000)             & Opt. Ctp.   & V, B-V, U-B  & $F_{x} (\mu Jy)$  & P$_{orb}$(hr)   \\
Name(s)    & DEC(J2000)            &             & E$_{B-V}$    & (2-10 keV)        & P$_{pulse}$(s)  \\
type       & $l^{II}, b^{II}$      & [FC]        &              &                   &                 \\

\noalign{\smallskip} \hline \noalign{\smallskip}
1H 1715-321     & 17 18 47.4      &            &          & 28          &     \\
                & -32 10 40       &            &          &             &     \\
B               & 354.1, +3.1     & [1277]     &          & [950, 1685] &     \\
\\
\hline
\end{tabular}
\end{table*}

\clearpage \begin{table*}[t]
\begin{tabular}{p{3.8cm}p{1.9cm}p{2.3cm}p{3.5cm}p{2.0cm}p{2.0cm}}
\noalign{\smallskip}
\multicolumn{6}{p{17.6cm}}{Table 1.  (continued) }\\
\hline \noalign{\smallskip}
Source     & RA(J2000)         & Opt. Ctp.  & V, B-V, U-B  & $F_{x} (\mu Jy)$  & P$_{orb}$(hr)    \\
Name(s)    & DEC(J2000)        &            & E$_{B-V}$    & (2-10 keV)        & P$_{pulse}$(s)   \\
type       & $l^{II}, b^{II}$  & [FC]       &              &                   &                  \\

 \noalign{\smallskip}\hline \noalign{\smallskip}
IGR J17269-4737  & 17 26 49.28  & star    &   & 83         &  \\
XTE J1727-476    & -47 38 24.9  &         &   & (2-12 keV) &  \\
T                & 342.2, -6.9  & [931]   &   & [883]      &  \\
\\
\hline
\end{tabular}
\end{table*}

\clearpage \begin{table*}[t]
\begin{tabular}{p{3.8cm}p{1.9cm}p{2.3cm}p{3.5cm}p{2.0cm}p{2.0cm}}
\noalign{\smallskip}
\multicolumn{6}{p{17.6cm}}{Table 1.  (continued) }\\
\hline \noalign{\smallskip}
Source     & RA(J2000)        & Opt. Ctp. & V, B-V, U-B & $F_{x} (\mu Jy)$  & P$_{orb}$(hr)  \\
Name(s)    & DEC(J2000)       &           & E$_{B-V}$   & (2-10 keV)        & P$_{pulse}$(s) \\
type       & $l^{II}, b^{II}$ & [FC]      &             &                   &                \\

\noalign{\smallskip} \hline \noalign{\smallskip}
MXB 1730-335    & 17 33 24.61       & Lil 1    &              & 200     &      \\
Rapid Burster   & -33 23 19.8       &          & 3.0          &         &      \\
T, G, B, D, R   & 354.8, -0.16      & [662]    & [442, 1183]  & [465]   &      \\
\\
\hline

\end{tabular}
\end{table*}

\clearpage \begin{table*}[t]
\begin{tabular}{p{3.8cm}p{1.9cm}p{2.3cm}p{3.5cm}p{2.0cm}p{2.0cm}}
\noalign{\smallskip}
\multicolumn{6}{p{17.6cm}}{Table 1.  (continued) }\\
\hline \noalign{\smallskip}
Source     & RA(J2000)         & Opt. Ctp.  & V, B-V, U-B  & $F_{x} (\mu Jy)$  & P$_{orb}$(hr)    \\
Name(s)    & DEC(J2000)        &            & E$_{B-V}$    & (2-10 keV)        & P$_{pulse}$(s)   \\
type       & $l^{II}, b^{II}$  & [FC]       &              &                   &                  \\

\noalign{\smallskip} \hline \noalign{\smallskip}
GRS 1737-31  & 17 40 09     &              &             & 26           &        \\
             & -31 02.4     &              &             &              &        \\
T            & 357.3, +0.58 &              &             & [973]        &        \\
\\
\hline

\end{tabular}
\end{table*}

\clearpage \begin{table*}[t]
\begin{tabular}{p{3.8cm}p{1.9cm}p{2.3cm}p{3.5cm}p{2.0cm}p{2.0cm}}
\noalign{\smallskip}
\multicolumn{6}{p{17.6cm}}{Table 1.  (continued) }\\
\hline \noalign{\smallskip}
Source     & RA(J2000)         & Opt. Ctp.  & V, B-V, U-B  & $F_{x} (\mu Jy)$  & P$_{orb}$(hr)    \\
Name(s)    & DEC(J2000)        &            & E$_{B-V}$    & (2-10 keV)        & P$_{pulse}$(s)   \\
type       & $l^{II}, b^{II}$  & [FC]       &              &                   &                  \\

\noalign{\smallskip} \hline \noalign{\smallskip}
GRS 1741.9-2853  & 17 45 02.33     &            &        & 7           &         \\
AX J1745.0-2855  & -28 54 49.7     &            &        &             & 0.0017    \\
T, B             & 356.0, +0.13    & [222]      &        & [1501]      &         \\
\\
\hline

\end{tabular}
\end{table*}

\clearpage \begin{table*}[t]
\begin{tabular}{p{3.8cm}p{1.9cm}p{2.3cm}p{3.5cm}p{2.0cm}p{2.0cm}}
\noalign{\smallskip}
\multicolumn{6}{p{17.6cm}}{Table 1.  (continued) }\\
\hline \noalign{\smallskip}
Source     & RA(J2000)         & Opt. Ctp.  & V, B-V, U-B  & $F_{x} (\mu Jy)$  & P$_{orb}$(hr)    \\
Name(s)    & DEC(J2000)        &            & E$_{B-V}$    & (2-10 keV)        & P$_{pulse}$(s)   \\
type       & $l^{II}, b^{II}$  & [FC]       &              &                   &                  \\

\noalign{\smallskip} \hline \noalign{\smallskip}
1E 1742.8-2853 & 17 45 59.5   &                &               & 0.2             &      \\
               & -28 55 01    &                &               & (0.9-4.0 keV)   &      \\
               & 0.06, -0.063 & [1689]         &               & [1689]          &      \\
\\
\hline

\end{tabular}
\end{table*}

\clearpage \begin{table*}[t]
\begin{tabular}{p{3.8cm}p{1.9cm}p{2.3cm}p{3.5cm}p{2.0cm}p{2.0cm}}
\noalign{\smallskip}
\multicolumn{6}{p{17.6cm}}{Table 1.  (continued) }\\
\hline \noalign{\smallskip}
Source     & RA(J2000)         & Opt. Ctp.  & V, B-V, U-B  & $F_{x} (\mu Jy)$  & P$_{orb}$(hr)    \\
Name(s)    & DEC(J2000)        &            & E$_{B-V}$    & (2-10 keV)        & P$_{pulse}$(s)   \\
type       & $l^{II}, b^{II}$  & [FC]       &              &                   &                  \\

\noalign{\smallskip} \hline \noalign{\smallskip}
SLX 1744-299    & 17 47 25.9      &            &               & 6             &       \\
AX J1747.4-3000 & -29 59 57.8     &            & $>$7.5:       &               &       \\
B               & 359.3, -0.89    &            & [1415]        & [1415]        &       \\
\\
\hline
\end{tabular}
\end{table*}

\clearpage \begin{table*}[t]
\begin{tabular}{p{3.8cm}p{1.9cm}p{2.3cm}p{3.5cm}p{2.0cm}p{2.0cm}}
\noalign{\smallskip}
\multicolumn{6}{p{17.6cm}}{Table 1.  (continued) }\\
\hline \noalign{\smallskip}
Source     & RA(J2000)         & Opt. Ctp.  & V, B-V, U-B  & $F_{x} (\mu Jy)$  & P$_{orb}$(hr)    \\
Name(s)    & DEC(J2000)        &            & E$_{B-V}$    & (2-10 keV)        & P$_{pulse}$(s)   \\
type       & $l^{II}, b^{II}$  & [FC]       &              &                   &                  \\

\noalign{\smallskip} \hline \noalign{\smallskip}
4U 1746-37      & 17 50 12.728     & $^*U1$       &              & 32            & 5.16     \\
                & -37 03 06.53     & NGC 6441     & 0.50         &               &         \\
G, B, A, D      & 353.5, -5.0      & [663]        & [1517]       & [1201, 1646]  & [44]  \\
\\
\hline
\end{tabular}
\end{table*}

\clearpage \begin{table*}[t]
\begin{tabular}{p{3.8cm}p{1.9cm}p{2.3cm}p{3.5cm}p{2.0cm}p{2.0cm}}
\noalign{\smallskip}
\multicolumn{6}{p{17.6cm}}{Table 1.  (continued) }\\
\hline \noalign{\smallskip}
Source     & RA(J2000)         & Opt. Ctp.  & V, B-V, U-B  & $F_{x} (\mu Jy)$  & P$_{orb}$(hr)    \\
Name(s)    & DEC(J2000)        &            & E$_{B-V}$    & (2-10 keV)        & P$_{pulse}$(s)   \\
type       & $l^{II}, b^{II}$  & [FC]       &              &                   &                  \\

\noalign{\smallskip} \hline \noalign{\smallskip}
SAX J1748.9-2021  & 17 48 52.163     & $^*V2$    &         & $<$0.1-150  &         \\
4U 1745-203       & -20 21 32.4      & NGC 6440  & 1.0     &             & 0.0024  \\
T, G, A, B        & 7.7, +3.8        & [1650]    & [1181]  & [773]       & [773]  \\
\\
\hline
\end{tabular}
\end{table*}

\clearpage \begin{table*}[t]
\begin{tabular}{p{3.8cm}p{1.9cm}p{2.3cm}p{3.5cm}p{2.0cm}p{2.0cm}}
\noalign{\smallskip}
\multicolumn{6}{p{17.6cm}}{Table 1.  (continued) }\\
\hline \noalign{\smallskip}
Source     & RA(J2000)         & Opt. Ctp.  & V, B-V, U-B  & $F_{x} (\mu Jy)$  & P$_{orb}$(hr)  \\
Name(s)    & DEC(J2000)        &            & E$_{B-V}$    & (2-10 keV)        & P$_{pulse}$(s) \\
type       & $l^{II}, b^{II}$  & [FC]       &              &                   &                \\

\noalign{\smallskip} \hline \noalign{\smallskip}
4U 1755-33      & 17 58 40.0     & V4134 Sgr    & 18.5, 0.7, - -   & 100       & 4.4         \\
                & -33 48 27      & $^*X$        & 0.5:             &           &             \\
T, U, D         & 357.2, -4.9    & [1666]       & [1002, 1141]     & [1685]    & [1002, 1708] \\
\\
\hline

\end{tabular}
\end{table*}

\clearpage \begin{table*}[t]
\begin{tabular}{p{3.8cm}p{1.9cm}p{2.3cm}p{3.5cm}p{2.0cm}p{2.0cm}}
\noalign{\smallskip}
\multicolumn{6}{p{17.6cm}}{Table 1.  (continued) }\\
\hline \noalign{\smallskip}
Source     & RA(J2000)         & Opt. Ctp.  & V, B-V, U-B  & $F_{x} (\mu Jy)$  & P$_{orb}$(hr)    \\
Name(s)    & DEC(J2000)        &            & E$_{B-V}$    & (2-10 keV)        & P$_{pulse}$(s)   \\
type       & $l^{II}, b^{II}$  & [FC]       &              &                   &                  \\

\noalign{\smallskip}\hline \noalign{\smallskip}
SAX J1806.5-2215 & 18 06 34     &         &   & 9        &  \\
                 & -22 15.1     &         &   &          &  \\
T, B             & 8.14, -0.70  & [286]   &   & [286]    &  \\
\\
\hline

\end{tabular}
\end{table*}

\clearpage \begin{table*}[t]
\begin{tabular}{p{3.8cm}p{1.9cm}p{2.3cm}p{3.5cm}p{2.0cm}p{2.0cm}}
\noalign{\smallskip}
\multicolumn{6}{p{17.6cm}}{Table 1.  (continued) }\\
\hline \noalign{\smallskip}
Source     & RA(J2000)         & Opt. Ctp.  & V, B-V, U-B  & $F_{x} (\mu Jy)$  & P$_{orb}$(hr)    \\
Name(s)    & DEC(J2000)        &            & E$_{B-V}$    & (2-10 keV)        & P$_{pulse}$(s)   \\
type       & $l^{II}, b^{II}$  & [FC]       &              &                   &                  \\

\noalign{\smallskip} \hline \noalign{\smallskip}
4U 1813-14   & 18 16 01.389    & $^*A$    &          & 1500        &        \\
GX 17+2      & -14 02 10.62    &          & $>$2     &             &        \\
B, Z, R      & 16.4, +1.3      & [352]    & [352]    & [462]       &        \\
\\
\hline
\end{tabular}
\end{table*}

\clearpage \begin{table*}[t]
\begin{tabular}{p{3.8cm}p{1.9cm}p{2.3cm}p{3.5cm}p{2.0cm}p{2.0cm}}
\noalign{\smallskip}
\multicolumn{6}{p{17.6cm}}{Table 1.  (continued) }\\
\hline \noalign{\smallskip}
Source     & RA(J2000)         & Opt. Ctp.  & V, B-V, U-B  & $F_{x} (\mu Jy)$  & P$_{orb}$(hr)  \\
Name(s)    & DEC(J2000)        &            & E$_{B-V}$    & (2-10 keV)        & P$_{pulse}$(s) \\
type       & $l^{II}, b^{II}$  & [FC]       &              &                   &                \\

\noalign{\smallskip} \hline \noalign{\smallskip}
4U 1823-00  & 18 25 22.02     & $^*X$       & 22, 1:, - -    & 25-62         & 3.2   \\
            & -00 00 43.0     &             & 1.3            &               &      \\
            & 29.9, +5.8      & [1683]      & [225, 607]     & [128]         &      \\
\\
\hline

\end{tabular}
\end{table*}

\clearpage \begin{table*}[t]
\begin{tabular}{p{3.8cm}p{1.9cm}p{2.3cm}p{3.5cm}p{2.0cm}p{2.0cm}}
\noalign{\smallskip}
\multicolumn{6}{p{17.6cm}}{Table 1.  (continued) }\\
\hline \noalign{\smallskip}
Source     & RA(J2000)         & Opt. Ctp.  & V, B-V, U-B  & $F_{x} (\mu Jy)$  & P$_{orb}$(hr)    \\
Name(s)    & DEC(J2000)        &            & E$_{B-V}$    & (2-10 keV)        & P$_{pulse}$(s)   \\
type       & $l^{II}, b^{II}$  & [FC]       &              &                   &                  \\

\noalign{\smallskip} \hline \noalign{\smallskip}
4U 1850-087  & 18 53 04.86  & $^*S$          & 21.0, 0.2, -0.8 & 10       & 0.343  \\
             & -08 42 20.4  & NGC 6712       & 0.33            &          &        \\
G, B, A, R?  & 25.4, -4.3   & [513, 657]     & [27, 861]       & [767]    & [657]  \\
\\
\hline

\end{tabular}
\end{table*}

\clearpage \begin{table*}[t]
\begin{tabular}{p{3.8cm}p{1.9cm}p{2.3cm}p{3.5cm}p{2.0cm}p{2.0cm}}
\noalign{\smallskip}
\multicolumn{6}{p{17.6cm}}{Table 1.  (continued) }\\
\hline \noalign{\smallskip}
Source     & RA(J2000)         & Opt. Ctp.  & V, B-V, U-B  & $F_{x} (\mu Jy)$  & P$_{orb}$(hr)    \\
Name(s)    & DEC(J2000)        &            & E$_{B-V}$    & (2-10 keV)        & P$_{pulse}$(s)   \\
type       & $l^{II}, b^{II}$  & [FC]       &              &                   &                  \\

\noalign{\smallskip}\hline \noalign{\smallskip}
XTE J1908+094 & 19 08 53.077 & star   &  & 100    &  \\
              & +09 23 04.90 &        &  &        &  \\
T, R          & 43.3, 0.43   & [215]  &  & [498]  &  \\
\\
\hline
\end{tabular}
\end{table*}

\clearpage \begin{table*}[t]
\begin{tabular}{p{3.8cm}p{1.9cm}p{2.3cm}p{3.5cm}p{2.0cm}p{2.0cm}}
\noalign{\smallskip}
\multicolumn{6}{p{17.6cm}}{Table 1.  (continued) }\\
\hline \noalign{\smallskip}
Source     & RA(J2000)         & Opt. Ctp.  & V, B-V, U-B  & $F_{x} (\mu Jy)$  & P$_{orb}$(hr)  \\
Name(s)    & DEC(J2000)        &            & E$_{B-V}$    & (2-10 keV)        & P$_{pulse}$(s) \\
type       & $l^{II}, b^{II}$  & [FC]       &              &                   &                \\

\noalign{\smallskip}\hline \noalign{\smallskip}
4U 1957+11  & 19 59 24.2   & V1408 Aql    & 18.7, 0.3, -0.6  & 30      & 9.33     \\
            & +11 42 32.4  &              & 0.4              &         &          \\
U           & 51.3, -9.3   & [372, 513]   & [947]            & [1685]  & [1543]   \\
\\
\hline
\end{tabular}
\end{table*}

\clearpage \begin{table*}[t]
\begin{tabular}{p{3.8cm}p{1.9cm}p{2.3cm}p{3.5cm}p{2.0cm}p{2.0cm}}
\noalign{\smallskip}
\multicolumn{6}{p{17.6cm}}{Table 1.  (continued) }\\
\hline \noalign{\smallskip}
Source     & RA(J2000)         & Opt. Ctp.  & V, B-V, U-B  & $F_{x} (\mu Jy)$  & P$_{orb}$(hr)  \\
Name(s)    & DEC(J2000)        &            & E$_{B-V}$    & (2-10 keV)        & P$_{pulse}$(s) \\
type       & $l^{II}, b^{II}$  & [FC]       &              &                   &                \\

\noalign{\smallskip}\hline \noalign{\smallskip}
CXO J212958.1+121002  & 21 29 58.06  & M15    & FUV$\sim$17   & 5           & 0.376  \\
M15-X2                & +12 10 02.6  & star   & 0.1           & (0.5-7 keV) &        \\
G, B                  & 65.0, -27.3  & [357]  & [583]         & [1700]      & [357] \\
\\
\hline

\end{tabular}
\end{table*}

\end{document}